\documentstyle[epsf]{article}

\def\etal{{\it et al.}}
\def\ie{{\it i.e.}}
\def\apriori{{\it a priori}}

\def\epem{e$^+$e$^-\;$}
\def\oalp{\relax\ifmmode O(\alpha_s)\else{{O($\alpha_s$)}}\fi}
\def\oalpsq{\relax\ifmmode O(\alpha_s^2)
           \else{{O($\alpha_s^2$)}}\fi}
\def\oalpc{\relax\ifmmode O(\alpha_s^3)
           \else{{O($\alpha_s^3$)}}\fi}

\begin{document}

\newcommand{\qq}{{${\rm q}\bar{\rm q}\;$}}
\newcommand{\qqg}{{${\rm q}\bar{\rm q}{\rm g}\;$}}

\def\d{{\rm d}}
\def\Im{{\rm Im\ }}
\def\Re{{\rm Re\ }}

\def\dth{\d\theta}
\def\dchi{\d\chi}
\def\eg{{\it e.g.}} 

\def \sup{^{\vphantom{2}}}

\def\sigmaU{\sigma\sup_{\rm U}}
\def\sigmaL{\sigma\sup_{\rm L}}
\def\sigmaT{\sigma\sup_{\rm T}}
\def\sigmaI{\sigma\sup_{\rm I}}
\def\sigmaA{\sigma\sup_{\rm A}}
\def\sigmaP{\sigma\sup_{\rm P}}

\def\hatsigmaU{\hat\sigma\sup_{\rm U}}
\def\hatsigmaL{\hat\sigma\sup_{\rm L}}
\def\hatsigmaT{\hat\sigma\sup_{\rm T}}
\def\hatsigmaI{\hat\sigma\sup_{\rm I}}
\def\hatsigmaA{\hat\sigma\sup_{\rm A}}
\def\hatsigmaP{\hat\sigma\sup_{\rm P}}

\def\sigmahatsum{\hatsigmaU+\hatsigmaL}

\def\tilAFB{\tilde A_{\rm FB}}

\def\vecone{{\mbox{\bf 1}}}
\def\vecthr{{\mbox{\bf 3}}}
\def\vecq{{\mbox{\bf q}}}
\def\vecqb{\bar{\mbox{\bf q}}}
\def\vecg{{\mbox{\bf g}}}
\def\vece{{\mbox{\bf e}}}

\def\FU{F\sup_{\rm U}}
\def\FL{F\sup_{\rm L}}
\def\FT{F\sup_{\rm T}}
\def\FI{F\sup_{\rm I}}
\def\FA{F\sup_{\rm A}}
\def\FP{F\sup_{\rm P}}

\def\cg{c_{\rm g}}
\def\sg{s_{\rm g}}
\def\thW{\theta_W}
\renewcommand{\thefootnote}{\fnsymbol{footnote}}
\phantom{MMM}
\begin{flushright}
\vspace{-10mm}
DESY 97-244\\ 
{\tt hep-ph/9712312} \\[1.7ex]
\end{flushright}

\begin{center}
{\Large \bf EVENT ORIENTATION IN e$^+$e$^-$ ANNIHILATION\footnote{To 
appear in:
{\it Proceedings of XIIth International Workshop: High Energy
Physics and Quantum Field Theory}, Samara, Russia,
September 4--10, 1997.}
}
\vspace{4mm}

P.N. Burrows\footnote{email: burrows@slac.stanford.edu}\\
Laboratory for Nuclear Science,
         Massachusetts Institute of Technology\\
  Cambridge, MA 02139, USA\\ 

and \\

P. Osland\footnote{email: per.osland@fi.uib.no} \\
Department of Physics, University of Bergen, All\'egt.~55,
N-5007 Bergen, Norway, and \\
Deutsches Elektronen-Synchrotron DESY, D-22603 Hamburg, Germany
\end{center}

\begin{abstract}
We review the orientation of \epem $\rightarrow$ \qqg events
in terms of the polar and azimuthal angles of the
event plane w.r.t.\ the electron beam direction.
The asymmetry of the azimuthal-angle distribution is,
like the left-right forward-backward
polar-angle asymmetry, sensitive to parity-violating
effects in three-jet events;
these are presently being explored experimentally.
We present these observables at \oalp\ in perturbative
QCD and discuss their dependence on longitudinal beam polarisation
and c.m.\ energy.
A moments analysis in terms of the orientation angles allows a 
more detailed test of QCD by isolating the 
independent helicity cross-sections.
\end{abstract}
%
\section{Introduction}
%
\setcounter{footnote}{0}
\renewcommand{\thefootnote}{\arabic{footnote}}

In \epem annihilation, events containing three distinct jets 
of hadrons were first observed at the PETRA storage ring in 1979 
\cite{PETRA}.
Such events were interpreted in terms of the fundamental
process \epem $\rightarrow$ \qqg, 
providing direct evidence for the existence of the gluon,
the vector boson of  QCD \cite{QCD}.
Subsequent studies of the properties of such events \cite{WU}
have confirmed this interpretation. 

Many jet observables have been explored experimentally in \epem
annihilations, yielding information on QCD as well as on the electroweak
theory.
We here consider the orientation of the \qqg plane 
or `event plane' \cite{BurrOsl}
in terms of the angles $\theta$ and $\chi$, 
where $\theta$ is the polar angle of
the quark direction with respect to the electron beam, and $\chi$ is
the azimuthal orientation angle of the event plane with respect to the
quark-electron plane, such that (Fig.~1):
\begin{equation}
\cos\chi =  \frac{\vecq \times \vecg}{
| \vecq \times \vecg|}
\cdot \frac{\vecq \times \vece^{-}}{
| \vecq \times \vece^{-}|}.
\label{eqchi}
\end{equation}
The polar angle can also be defined in two-jet
events of the type \epem $\rightarrow$ \qq, in which case the distribution
in $\theta$ is determined in the electroweak theory
\cite{EW},
and displays a c.m.\ energy-dependent 
forward-backward asymmetry which has been observed in many
experiments \cite{AFBREV,BLONDEL}.
For \epem annihilation at the $Z^0$ resonance, the polar-angle 
asymmetry is large only if one, or both, of the beams are longitudinally 
polarised, as at SLC/SLD \cite{SLDAFB}.
The azimuthal angle $\chi$ is, of course, undefined in \qq events,
but in \qqg events, it also displays an
asymmetry which can be large at the $Z^0$ resonance in
the case of highly polarised electrons.
This azimuthal-angle distribution is currently being investigated
experimentally \cite{SLD-HamJer}.

%
\section{The \epem $\rightarrow{\bf q}\bar{\bf q}{\bf g}\;$
Differential Cross-Section}
%
Let $\vecq$, $\vecqb$, and $\vecg$ denote the quark, antiquark
and gluon momenta respectively (see Fig.~1) and
$x$, $\bar x$ and $x_g$ be the scaled energies (for simplicity,
we consider the limit of massless quarks):
\begin{equation}
|\vecq\,|=x\sqrt{s}, \qquad |\vecqb\,|=\bar x\sqrt{s}, \qquad 
|\vecg\,|=x_g\sqrt{s},
\end{equation}
with  $x+\bar x+x_g=2.$
\begin{figure}[htb]
\begin{center}
\setlength{\unitlength}{1cm}
\begin{picture}(10,6.5)
\put(1.5,-1.0)
{\mbox{\epsfysize=8.0cm\epsffile{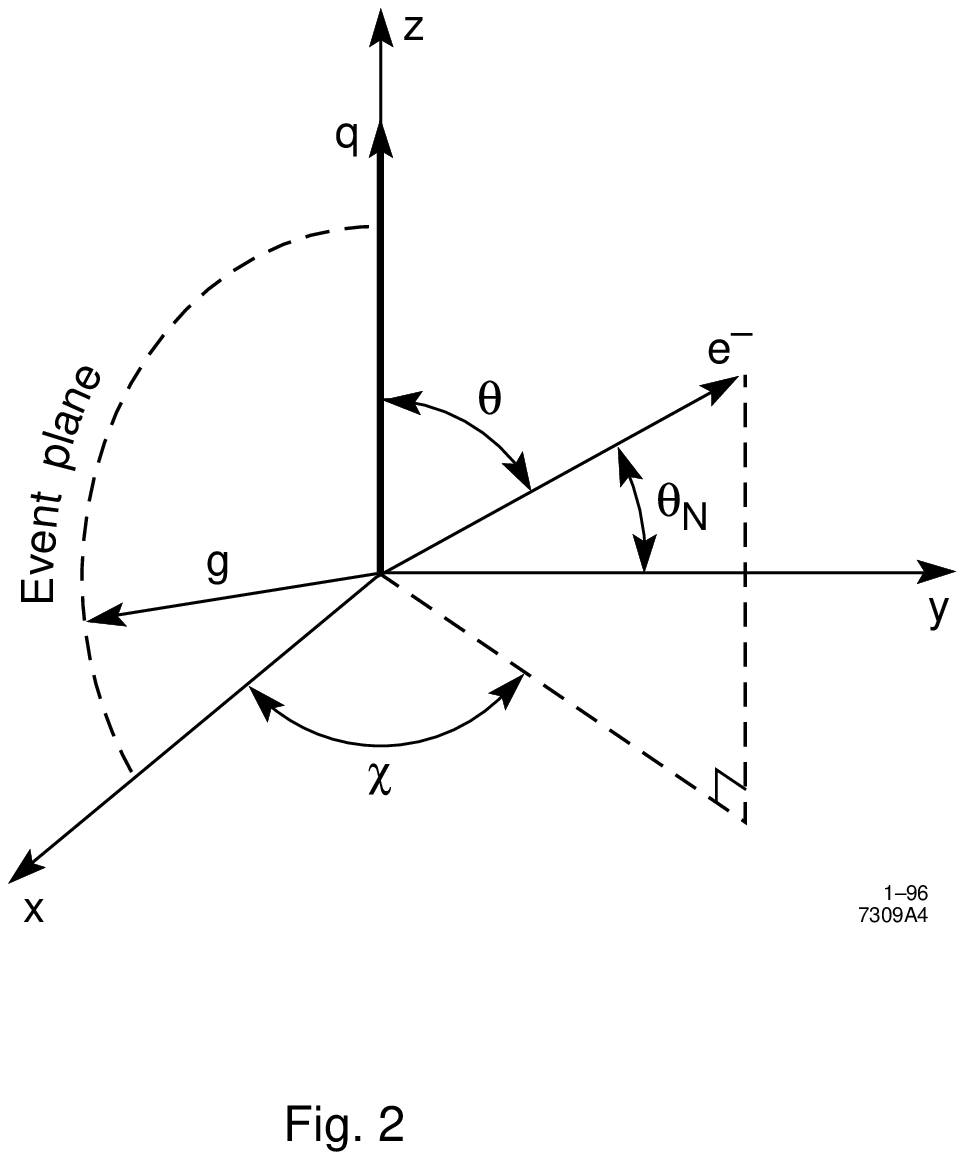}}}
\end{picture}
\begin{picture}(10,3.0)
\put(2.5,1.5)
{\mbox{\epsfysize=1.0cm\epsffile{whiteout.eps}}}
\end{picture}
\vspace*{-35mm}
%
%
\caption{Definition of the angles $\theta$ and $\chi$.}
\end{center}
\end{figure}
Allowing for longitudinal beam polarisation,
the fully-differential three-jet cross-section can then at the
tree-level, and for massless quarks, be expressed as \cite{OOO}:
\begin{eqnarray}
2\pi{\d^4\sigma \over \d(\cos\theta)\d\chi\d x\d\bar x} 
&=& \biggl[
\frac{3}{8}(1+\cos^2\theta)\, {\d^2\sigmaU\over \d x\d\bar x}
+\frac{3}{4}\sin^2\theta\, {\d^2\sigmaL\over \d x\d\bar x} \nonumber \\
& &+\frac{3}{4}\sin^2\theta\cos2\chi\, {\d^2\sigmaT\over \d x\d\bar x}
+{3\over2\sqrt{2}}\sin2\theta\cos\chi\, {\d^2\sigmaI\over \d x\d\bar x}
\biggr] h_f^{(1)}(s) \nonumber \\
& &+ \biggl[
\frac{3}{4}\cos\theta\, {\d^2\sigmaP\over \d x\d\bar x}\;
-\;{3\over\sqrt{2}}\sin\theta\cos\chi\, {\d^2\sigmaA\over \d x\d\bar x}
\biggr]h_f^{(2)}(s),
\label{eqfour}
\end{eqnarray}
where at lowest order in the electroweak theory
the dependences on flavour and beam polarisation
are very transparent, and given by the functions:
\begin{eqnarray}
h_f^{(1)}(s) &=&
Q_f^2\Xi -2Q_f\Re f(s)(v\Xi-a\xi)v_f 
+|f(s)|^2[(v^2+a^2)\Xi-2va\xi](v_f^2+a_f^2), \label{eqhfone} \\
h_f^{(2)}(s) &=&
-2Q_f\Re f(s)(a\Xi-v\xi)a_f 
+|f(s)|^2[-(v^2+a^2)\xi+2va\Xi]2v_fa_f, \label{eqhftwo}
\end{eqnarray}
with
\begin{equation}
f(s)={1\over 4\sin^22\thW}\; {s\over s-M_Z^2+iM_Z\Gamma_Z^{\rm tot}},
\end{equation}
where $Q_f$ is the charge of quark flavour $f$. Furthermore,
$v$, $a$ ($v_f$, $a_f$) are the vector and axial vector couplings of
the $Z^0$ to the electron (quark of flavour $f$), respectively.
The longitudinal beam polarisations enter through the coefficients
\begin{equation}
\Xi=1-P_-^{\|}P_+^{\|}, \qquad \xi=P_-^{\|}-P_+^{\|}, 
\end{equation}
where $P_-^{\|}$ and $P_+^{\|}$ are the longitudinal polarisations
of the electron and positron beams.

Several important points concerning eq.~(\ref{eqfour}) should be noted.
The cross-section can be written 
as a sum of 6 terms, each of which may be factorised
into three contributions: the first factor
is a simple trigonometric function
of the polar and azimuthal orientation angles
$\theta$ and $\chi$, and the second, 
$\d^2\sigma_i/\d x \d \bar x$ ($i$=U, L, T, I, P, A),
is a function of the parton momentum fractions; 
these are determined by QCD and kinematics; 
the third factor, $h_f^{(1,2)}(s)$,
is a function containing the dependence on the fermion electroweak
couplings.
Hence, in each term there is factorisation both between 
the dynamical contributions of the QCD and electroweak sectors of the
Standard Model and between the orientation of the event plane and the
relative orientation of the jets within the plane.
One may exploit this property by defining moments in terms of
$\cos\theta$ and $\cos\chi$ in order to isolate the different terms
\cite{BurrOsl}.
The $\sigma_i$ are often referred to in the literature
as {\it helicity cross-sections},
and the form of eq.~(\ref{eqfour}), with
six terms, each containing one of the 
independent helicity cross-sections, has been shown \cite{KS}
to be valid for massless partons up to \oalpsq\ in perturbative QCD.

When quark masses are introduced, each of the first four terms 
in eq.~(\ref{eqfour}) (proportional to $h_f^{(1)}(s)$) 
gets an additional contribution proportional to 
$\tilde h_f^{(1)}(s)\,\d^2\tilde\sigma_i$, where
\begin{equation}
\tilde h_f^{(1)}(s) 
=|f(s)|^2[(v^2+a^2)\Xi-2va\xi]a_f^2, 
\end{equation}
and where $\d^2\tilde\sigma_i$ is proportional to the square of 
the quark mass, $m_f^2$.

At \oalp\ in perturbative QCD \cite{OOO,Laermann}:
\begin{equation}
{\d^2\sigma_i\over\d x \d \bar x} =
{\displaystyle{\tilde\sigma\over(1-x)(1-\bar x)}}\, F_i,
\label{eqthree}
\end{equation}
where
\begin{equation}
\tilde\sigma={4\pi\alpha^2\alpha_s(s)\over3\pi s}.
\end{equation}
For the coordinate system of Fig.~1 the 
$\FU$, $\FL$, $\FT$, $\FI$, $\FA$ and $\FP$ are given in \cite{OOO},
entirely in terms of `internal' angles of the event.
Since the quark and antiquark tend to have a small
acollinearity angle, $\FL$, $\FT$, $\FI$ and $\FA$ will typically 
be small compared with $\FU$ and $\FP$.

The results presented here are at \oalp.
Corresponding results at \oalpsq, for massless quarks, could in
principle be derived from the helicity cross-section expressions given
in refs.~\cite{KS,SchuKor}.
For massive quarks the helicity cross-sections have
recently been calculated at \oalpsq\ in ref.~\cite{Aachen}; 
in this case there are
three additional cross-sections, corresponding to three new angular
dependences:
\begin{equation}
\sin2\theta\sin\chi, \qquad \sin^2\theta\sin2\chi, \qquad 
\mbox{and} \qquad \sin\theta\sin\chi.
\end{equation}
These terms are generated by absorptive parts in the scattering 
amplitude; they vanish in the massless limit even
at the one-loop order. Thus, they are quite small.
The last term, $\sin\theta\sin\chi$ can also be written as
$\pm\cos\omega$, where $\omega$ is the angle between the normal
to the event plane and the beam direction.
%
\section{Polar- and Azimuthal-Angle Distributions}
%
We now discuss the singly-differential cross-sections in terms of
$\cos\theta$ or $\chi$. 
Consider integrating eq.~(\ref{eqfour}) first over $x$ and
$\bar{x}$, with the integration domain given by some standard
jet resolution criterion $y_c$ \cite{SB} and using the notation:
\begin{equation}
\hat\sigma_i  \equiv 
\int_{y_c}\d x\int\d\bar x
{\d^2\sigma_i \over \d x\d\bar x}. 
\label{eqycut}
\end{equation}
Integrating over $\chi$ we then obtain:
\begin{equation}     {\d  \sigma \over \d(\cos\theta)}
 = \left( \frac{3}{8}(1+\cos^2\theta)\, \hat \sigmaU
+\frac{3}{4}\sin^2\theta\,  \hat\sigmaL \right) h_f^{(1)}(s)
 +\frac{3}{4}\cos\theta\,  \hat\sigmaP \, h_f^{(2)}(s),
\label{eqathe}
\end{equation}
where the term containing $\hat\sigmaP$ represents the well-known quark
forward-backward asymmetry resulting from parity violation in the weak
interaction, but for the three-jet case.
Similarly, by integrating over $\cos\theta$ we obtain:
\begin{equation}
2\pi\, {\d \sigma \over \d\chi}
= \left( \hat \sigmaU + \hat \sigmaL
+\cos 2\chi\, \hat \sigmaT \right) h_f^{(1)}(s)
-{3\pi\over2\sqrt{2}}\, \cos\chi\, \hat\sigmaA \, h_f^{(2)}(s),
\label{eqachi}
\end{equation}
where the term containing $\hat\sigmaA$ represents an azimuthal,
parity-odd asymmetry analogous to the last term in
eq.~(\ref{eqathe}) but owing its existence to the radiation of the gluon.

For the case of longitudinally-polarised electrons and unpolarised 
positrons the dependences of these singly-differential distributions
on the beam polarisation and c.m.\
energy are illustrated in Figs.~2(a,b) and 3(a,b) respectively.
\begin{figure}[htb]
\begin{center}
\setlength{\unitlength}{1cm}
\begin{picture}(10,4.5)
\put(0.,-2.0)
{\mbox{\epsfysize=7.0cm\epsffile{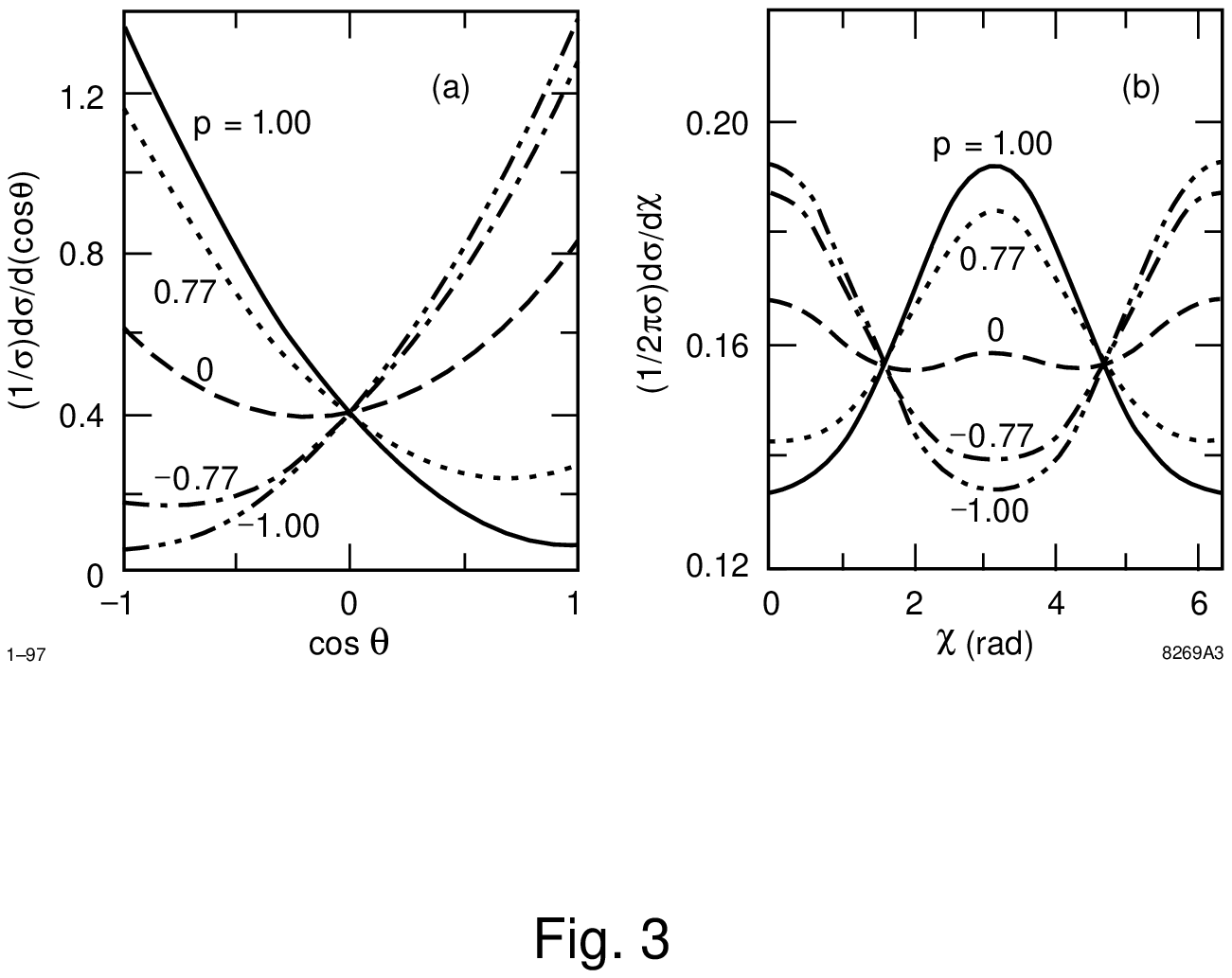}}}
\put(3.,-2.0)
{\mbox{\epsfysize=1.0cm\epsffile{whiteout.eps}}}
\end{picture}
\vspace*{-2mm}
%
%
\caption{Angular orientation of
the event plane for down-type quarks and $y_c=0.02$.
Distribution of (a) $\cos\theta$ and (b) $\chi$
at $\sqrt{s}=M_Z$, for 5 values of $p$.}
\end{center}
\end{figure}
We present the $\sigma_i$ at \oalp\ as in Ref.~\cite{OOO}.
Fig.~2a shows the distribution in
$\cos\theta$ at $\sqrt{s}$ = $M_Z$ for down-type
quarks, $y_c=0.02$ and electron
longitudinal polarisation $p$ = $+1$, 0 and $-1$.
(We refer to
positive (negative) polarisation as right- (left-) handed respectively.)
The recent SLC/SLD case of $p$ = $\pm0.77$
\cite{SLDninetysix}
is also indicated. The quark polar-angle forward-backward
asymmetry is large for high beam polarisation, and its sign
changes with the sign of the polarisation.
The less familiar azimuthal-angle distribution is shown
in Fig.~2b for the same cases as in Fig.~2a;
the distribution is symmetric about $\chi=\pi$.
The phase change of the
$\chi$ distribution when the beam polarisation sign is changed is a
reflection of the sign reversal of the forward-backward asymmetry in
$\cos\theta$. Qualitatively similar results are
obtained for up-type quarks, and for other values of $y_c$. 

We illustrate the energy dependence of the $\cos\theta$- and 
$\chi$-distributions in Figs.~3a and 3b, respectively,
for down-type quarks at fixed electron polarisation $p$ = $+1$, 
with results at $\sqrt{s}$ = 35, 60, 91 and 200 GeV,
corresponding to e$^+$e$^-$ annihilation at the PETRA, TRISTAN, SLC/LEP and 
LEP2 collider energies.
The variation with energy is due to the varying relative
contribution of $\gamma$ and $Z^0$ exchange 
in the e$^+$e$^-$ annihilation process.
Results are also shown for a possible high-energy collider operating
with polarised electrons at $\sqrt{s}$ = 500 GeV and 2 TeV.
If such a facility could be operated at lower energies, where, 
apart from $\sqrt{s}=91$~GeV (SLC), polarised beams were
not previously available, measurements in the same experiment
of the distributions shown in Figs.~3(a,b) would
provide a significant consistency check of the Standard Model.
\begin{figure}[htb]
\begin{center}
\setlength{\unitlength}{1cm}
\begin{picture}(10,4.5)
\put(0.,-2.0)
{\mbox{\epsfysize=7.0cm\epsffile{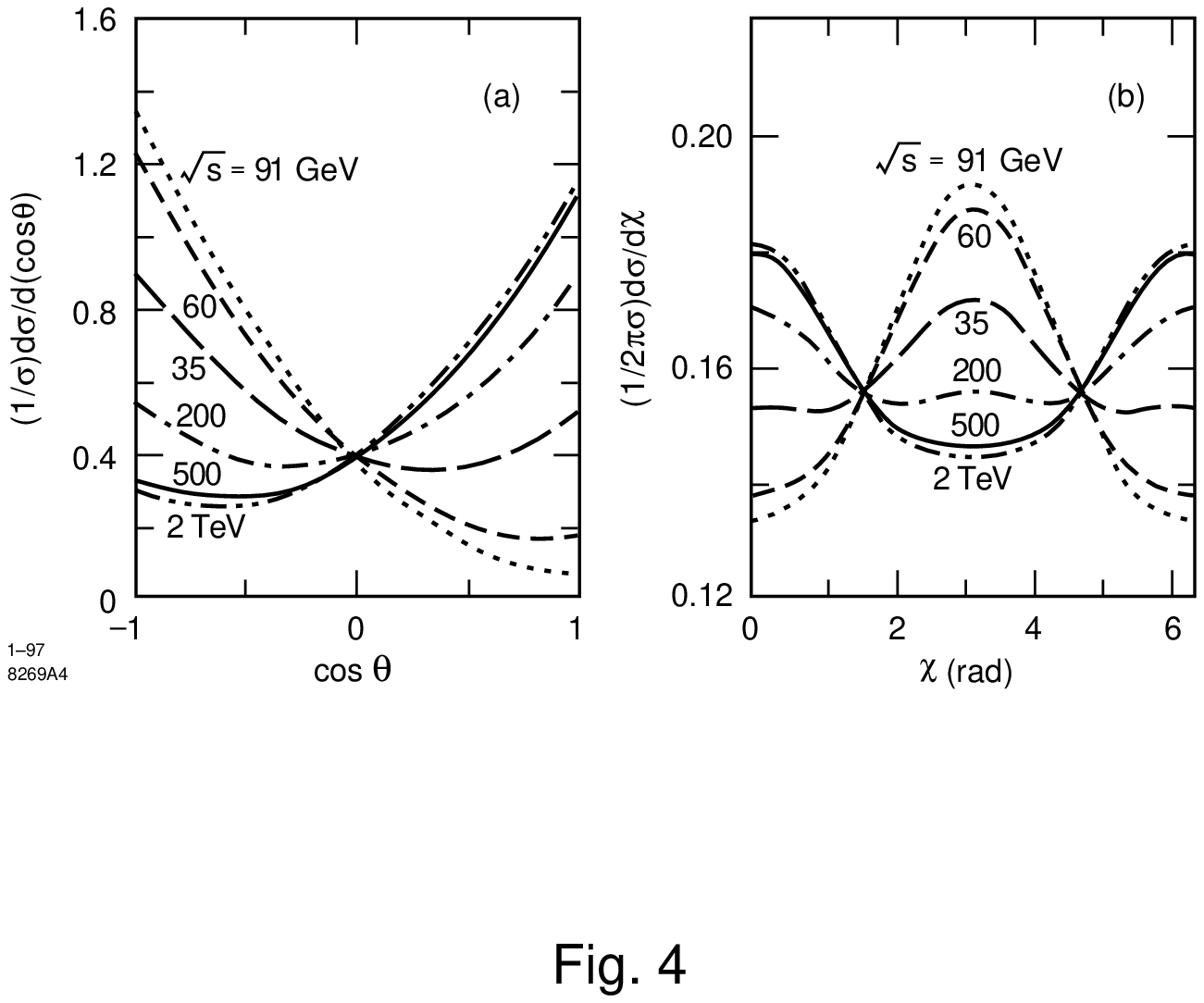}}}
\put(3.,-2.5)
{\mbox{\epsfysize=1.0cm\epsffile{whiteout.eps}}}
\end{picture}
\vspace*{-2mm}
%
%
\caption{Angular orientation of
the event plane for down-type quarks and $y_c=0.02$.
Distribution of (a) $\cos\theta$ and (b) $\chi$
for $p=+1$ at 6 values of $\sqrt{s}$.}
\end{center}
\end{figure}

%
\section{Polar- and Azimuthal-Angle Asymmetries}
%
By analogy with the {\it left-right forward-backward asymmetry}
of the polar-angle distribution:
\begin{equation}
\label{eqAFBtheta}
\tilAFB(|p|)|_{\cos\theta}
\equiv
\frac{
\int_0^1\,{\d\sigma^{\rm L}\over \d\cos\theta}\,\d\cos\theta
-
\int_{-1}^0\,
          {\d\sigma^{\rm L}\over \d\cos\theta}\,\d\cos\theta
-\left(
\int_0^1\,{\d\sigma^{\rm R}\over \d\cos\theta}\,\d\cos\theta
-
\int_{-1}^0\,
          {\d\sigma^{\rm R}\over \d\cos\theta}\,\d\cos\theta
\right)}
{
\int_0^1\,{\d\sigma^{\rm L}\over \d\cos\theta}\,\d\cos\theta
+
\int_{-1}^0\,
          {\d\sigma^{\rm L}\over \d\cos\theta}\,\d\cos\theta
+
\int_0^1\,{\d\sigma^{\rm R}\over \d\cos\theta}\,\d\cos\theta
+
\int_{-1}^0\,
          {\d\sigma^{\rm R}\over \d\cos\theta}\,\d\cos\theta
},
\end{equation}
it is natural to define a corresponding asymmetry of the azimuthal-angle
distribution \cite{BurrOsl}:
\begin{equation}
\tilde A(|p|)|_{\chi}\;
\equiv\;
{
\int_{{\pi\over 2}}^{\pi}
\,{\d\sigma^{\rm L}\over \d\chi}\,\d\chi
-
\int_0^{{\pi\over 2}}
\,{\d\sigma^{\rm L}\over \d\chi}\,\d\chi
-\left(
\int_{{\pi\over 2}}^{\pi}
\,{\d\sigma^{\rm R}\over \d\chi}\,\d\chi
-
\int_0^{{\pi\over 2}}
          {\d\sigma^{\rm R}\over \d\chi}\,\d\chi
\right)
\over{
\int_{{\pi\over 2}}^{\pi}
\,{\d\sigma^{\rm L}\over \d\chi}\,\d\chi
+
\int_0^{{\pi\over 2}}
\,{\d\sigma^{\rm L}\over \d\chi}\,\d\chi
+
\int_{{\pi\over 2}}^{\pi}
\,{\d\sigma^{\rm R}\over \d\chi}\,\d\chi
+
\int_0^{{\pi\over 2}}
\,{\d\sigma^{\rm R}\over \d\chi}\,\d\chi
}},
\label{eqAFBchi}
\end{equation}
where $\sigma^{\rm L,R}=\sigma(\mp|p|)$ is 
the \epem $\rightarrow$ \qqg cross-section
for a left- (L) or right- (R) handed electron beam of polarisation
magnitude $|p|$. 

For the case of e$^+$e$^-$ annihilation at the $Z^0$ resonance 
using electrons of
longitudinal polarisation $p$ and unpolarised positrons,
as at SLC, eqs.~(\ref{eqhfone}) and (\ref{eqhftwo}) reduce to the simple forms
\begin{eqnarray}
h_f^{(1)}(M_Z^2)  &=&
 |f(M_Z^2)|^2[(v^2+a^2)   -2vap  ](v_f^2+a_f^2), \nonumber \\
h_f^{(2)}(M_Z^2)  &=&
 |f(M_Z^2)|^2[-(v^2+a^2)p  +2va   ]2v_fa_f,  
\label{eqsimple}
\end{eqnarray}
and hence
\begin{equation}
\tilAFB(|p|)|_{\cos\theta}
= \frac{3}{4}\;|p|\;{\hat\sigmaP\over{\hat\sigmaU+\hat\sigmaL}}\;A_f, \qquad
\tilde A(|p|)|_{\chi}= {3\over\sqrt{2}}\;|p|\;
{\hat\sigmaA\over{\hat\sigmaU+\hat\sigmaL}}\;A_f,
\label{eqAFBboth}
\end{equation}
where we use the common notation $A_f \equiv 2v_f a_f/(v_f^2+a_f^2)$.
Whereas both asymmetries are directly proportional to the beam 
polarisation $|p|$ and the electroweak coupling $A_f$,
the $\cos\theta$ asymmetry is proportional to the
helicity cross-section $\hat\sigmaP$, and
the $\chi$ asymmetry to the helicity cross-section $\hat\sigmaA$.
Since the electroweak factor $A_f$ is predicted to a high degree of
accuracy by the Standard Model, and, in the case of b and c quarks,
has also been measured using predominantly \qq final states 
at SLC and LEP \cite{BLONDEL},
measurement of these asymmetries (\ref{eqAFBboth}) in \qqg events at
SLC/SLD would allow one to probe $\hat\sigmaP$ and $\hat\sigmaA$.
Preliminary results in this direction were reported recently
\cite{SLD-HamJer}.
Furthermore,
the ratio of the asymmetries is independent of both
polarisation and electroweak couplings and depends only on the ratio
of $\hat\sigmaP$ and $\hat\sigmaA$:
\begin{equation}
{\tilAFB(|p|)|_{\cos\theta}\over{\tilde A(|p|)|_{\chi}}}
= {\sqrt{2}\over 4}\;{\hat\sigmaP\over{\hat\sigmaA}}.
\label{eqratio}
\end{equation}
As a consequence of there being, up to \oalpsq\ in massless
perturbative QCD, only the six independent
helicity cross-sections given in eq.~(\ref{eqfour}),
the relations (\ref{eqAFBboth}) and (\ref{eqratio}) are valid 
up to the same order. 

We  show in Fig.~4a, at \oalp\ the ratios 
$\hat\sigmaP/(\hat\sigmaU+\hat\sigmaL)$ and
$\hat\sigmaA/(\hat\sigmaU+\hat\sigmaL)$ 
and their dependence on $y_c$; the dependence is weak.
For completeness we also show
$\hat\sigmaT/(\hat\sigmaU+\hat\sigmaL)$ and
$\hat\sigmaI/(\hat\sigmaU+\hat\sigmaL)$.
It would be worthwhile to investigate the size of higher-order
perturbative QCD contributions by evaluating these ratios at \oalpsq;
this should be possible using the matrix elements
described in Ref.~\cite{CG} or the results of \cite{Aachen}.
\begin{figure}[htb]
\begin{center}
\setlength{\unitlength}{1cm}
\begin{picture}(10,6.5)
\put(0.,-1.0)
{\mbox{\epsfysize=8.0cm\epsffile{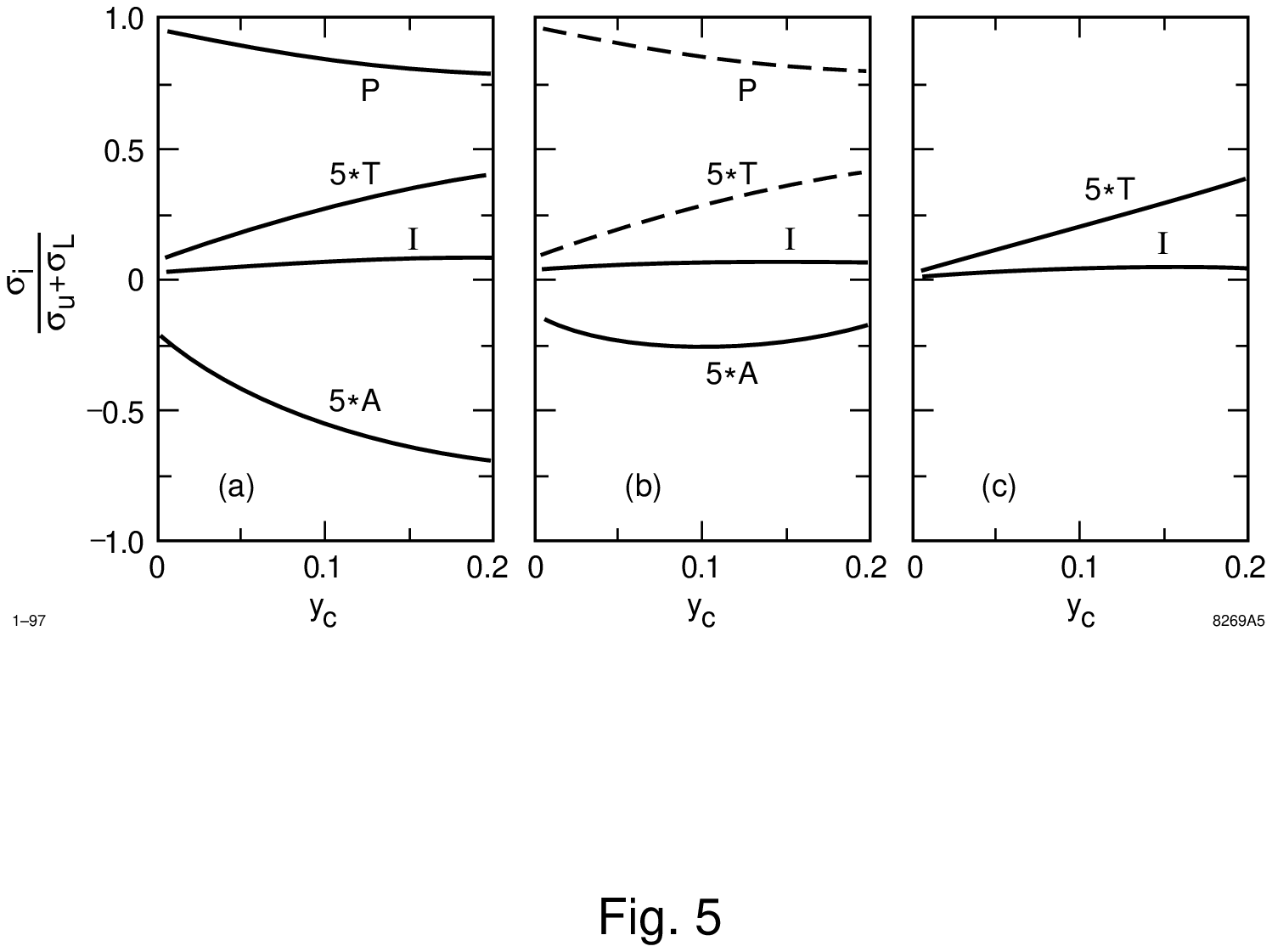}}}
\put(3.8,-1.0)
{\mbox{\epsfysize=1.0cm\epsffile{whiteout.eps}}}
\end{picture}
\vspace*{-15mm}
%
%
\caption{Helicity cross-section ratios
(see text) as functions of $y_c$;
(a) exclusive, (b) semi-inclusive and (c) fully-inclusive cases.
For the sake of clarity, the ratios are multiplied by
a factor of 5 where indicated.}
\end{center}
\end{figure}

It would be interesting to confront the theoretical predictions with
experimental measurements, taking into account mass effects and
higher-order QCD effects \cite{Aachen}.
Significant deviations of the data from the
predictions for the asymmetries, eqs.~(\ref{eqAFBboth}), 
would indicate anomalous
parity-violating contributions to the process 
\epem $\rightarrow$ \qqg. The ratio
of asymmetries, eq.~(\ref{eqratio}), is at lowest order independent of the
electroweak coupling factor $A_f$, and would help to unravel the
dynamical origin of any such effect.
%
\section{Inclusive Cross-Sections}
%
All of the preceding discussions have been based on the assumption
that the parton-type originator of jets is known, \ie\ that in 
3-jet events one
can identify which jet originated from the quark, antiquark and
gluon.
The definition of $\cos\theta$ requires that
the quark jet be known, whereas the definition of $\chi$
requires that both the quark jet and a second jet
origin be identified. It is
difficult from an experimental point-of-view to make such
exclusive
identification for jets of hadrons measured in a detector.
Quark and antiquark jets have been identified in predominantly 2-jet
events in \epem annihilation (see \eg\ \cite{SLDAFB}).
Currently, one has reached a single-hemisphere efficiency 
for b or $\bar{\rm b}$ quarks of the order of 70\%,
and a double-tagging efficiency of about 50\%.
Identification of both quark and antiquark jets in 
\epem $\rightarrow$ \qqg events is \apriori\ more
difficult due to the greater hadronic activity.

It is therefore useful to consider more inclusive quantities. 
Two possibilities are:
(1) {\it Semi-inclusive}: the quark jet is assumed to be identified,
and the {\it least energetic} jet in the event is taken to be the gluon
and is used to define the angle $\chi$ (eq.~(\ref{eqchi})).
While this assumption will be wrong part of the time,
it will be wrong by a calculable probability, and hence this can
in principle be fully corrected for.
(2) {\it Fully-inclusive}:
the jets are labelled only in terms of their energies,
$x_3\le x_2\le x_1$;
the polar angle $\theta$ is then defined by the angle of the fastest
jet w.r.t. the electron beam direction
and the azimuthal angle $\chi$ can be defined analogously to
eq.~(\ref{eqchi}) as: 
\begin{equation}
\cos\chi =  \frac{\vecone\times \vecthr}{
|\vecone\times\vecthr|}
\cdot \frac{\vecone \times \vece{}^-}{
|\vecone \times \vece{}^-|}
\label{eqThrchi}
\end{equation}

For the semi-inclusive case we show at \oalp\ the ratios 
$\hat\sigmaT/(\hat\sigmaU+\hat\sigmaL)$,
$\hat\sigmaI/(\hat\sigmaU+\hat\sigmaL)$,
$\hat\sigmaP/(\hat\sigmaU+\hat\sigmaL)$ and
$\hat\sigmaA/(\hat\sigmaU+\hat\sigmaL)$ 
in Fig.~4b.
Whereas $\hat\sigmaP$ and $\hat\sigmaT$ are unchanged relative to the
exclusive case, $\hat\sigmaI$ and $\hat\sigmaA$, which multiply
terms proportional to $\cos\chi$ in eq.~(\ref{eqfour}), are smaller
in magnitude because of the sometimes incorrect gluon-jet identification.
Though this implies that the parity-violating asymmetry
$\tilde A(|p|)|_{\chi}$ in eq.~(\ref{eqAFBboth}) is smaller
by a ($y_c$-dependent) factor of order 2,
it will in fact be easier to access experimentally because 
the semi-inclusive case requires only one of the quark- 
and antiquark-jets to be identified explicitly.

In the fully-inclusive case the terms $\sigmaA$ and
$\sigmaP$, which are odd under interchange of quark and antiquark jets,
cancel out; writing the cross-section in terms of thrust
\cite{THRUST}
one obtains at \oalp:
\begin{eqnarray}
2\pi{\d^3\sigma \over \d(\cos\theta)\d\chi\d T}
&=&\frac{3}{8}(1+\cos^2\theta)\, {\d\sigmaU\over \d T}
+\frac{3}{4}\sin^2\theta\, {\d\sigmaL\over \d T} \\
&&+\frac{3}{4}\sin^2\theta\cos2\chi\, {\d\sigmaT\over \d T}
+{3\over2\sqrt{2}}\sin2\theta\cos\chi\, {\d\sigmaI\over \d T},
\end{eqnarray}
where expressions for $\d\sigma_i/\d T$ can be found in 
ref.~\cite{OOO}.
Using the notation
\begin{equation}
\tilde\sigma_i \equiv \int_{y_c}{\d\sigma_i\over \d T}\,\d T,
\qquad \hbox{$i=$ U, L, T, I},
\end{equation}
we show at \oalp\ the ratios
$\tilde\sigmaT/(\tilde\sigmaU+\tilde\sigmaL)$ and
$\tilde\sigmaI/(\tilde\sigmaU+\tilde\sigmaL)$ in Fig.~4c.
Their magnitudes and dependences on $y_c$ differ relative to the
exclusive and semi-inclusive cases due to the redefinition of $\theta$
and $\chi$.
Distributions of $\cos\theta$ and $\chi$ in 
this case have already been measured 
and found to be in agreement with \oalp\ QCD calculations
\cite{TASSO,LEPone}.

Another fully-inclusive observable is the polar-angle $\omega$
of the normal to the event plane with respect to the beam direction.
The differential cross-section $\d\sigma/\d(\cos\omega)$
has been calculated at \oalpsq\ in massless perturbative QCD
\cite{THEBR},
and has been measured at $\sqrt{s}\simeq35$~GeV \cite{TASSO}
and $\sqrt{s}=91$~GeV \cite{LEPone}.
The effects of final-state interactions can induce a term linear
in $\cos\omega$ whose sign and magnitude depend on the electron
beam polarisation
\cite{Dixon};
experimental limits on such a term have been set using hadronic
$Z^0$ decays \cite{SLDthree}.
%
\section{Conclusions}
%
We have presented the orientation of 
\epem $\rightarrow$ \qqg events in terms of the
polar- ($\theta$) and azimuthal- ($\chi$) angle distributions. 
These distributions have been given at \oalp\ in perturbative QCD 
for massless quarks and their dependence on longitudinal electron-beam
polarisation and centre-of-mass energy has been illustrated.
The more complicated \oalpsq\ results are available for massless
quarks \cite{KS,SchuKor} and coming soon also for massive quarks 
\cite{Aachen}.
We have considered the left-right forward-backward asymmetry of the
$\cos\theta$ distribution and have presented a corresponding asymmetry of
the $\chi$ distribution. Parity-violating 3-jet observables of this kind
represent a new search-ground for anomalous contributions and
are presently being explored experimentally \cite{SLD-HamJer}.

For the case of \epem annihilation at the $Z^0$ resonance using 
longitudinally-polarised
electrons, the $\cos\theta$ asymmetry is proportional to
the QCD helicity cross-section $\hat\sigmaP$, and the $\chi$ asymmetry to
the helicity cross-section $\hat\sigmaA$;
these are now being measured using the
highly-polarised electron beam at SLC/SLD.
To lowest electroweak order the ratio of these asymmetries is
independent of electroweak couplings and the beam polarisation. 
These results are valid up to \oalpsq\ in QCD perturbation theory for
massless quarks.
At \oalp\ the dependence of $\hat\sigmaP$ and $\hat\sigmaA$ 
on the jet resolution parameter $y_c$ is found to be weak.
Higher-order perturbative QCD contributions, as well as quark mass effects, 
should be included before making a detailed
comparison of these predictions with data.
Even the extraction of $\hat\sigmaU$, $\hat\sigmaL$, $\hat\sigmaT$ and
$\hat\sigmaI$, which does not require quark and antiquark jet 
identification, represents a detailed test of QCD, beyond what has so
far been studied.

\medskip
It is a pleasure to thank the Organisers of the Samara Workshop,
in particular Professor V. Savrin,
for creating a very stimulating and pleasant atmosphere during the meeting.
This research has been supported by U.S.\ Department of Energy
Cooperative Agreement DE-FC02-94ER40818
and by the Research Council of Norway.
P.O. would like to acknowledge the kind hospitality of the Theory Group 
at DESY.

\end{document}